\documentstyle[11pt,NVO,twoside]{article}
\def\etal{\it et al. \rm }
\begin{document}

\title{NVO and the LSB Universe}

\author{James M. Schombert}
\affil{Department of Physics, University of Oregon, Eugene}

\begin{abstract}

There is tremendous scientific potential in a National Virtual
Observatory, particularly for projects that need to mine large databases
for rare or unusual objects.  However, the NVO will also make an impact on any
project, large or small, the requires a mixture of datasets to explore a
wide range of astrophysical phenomenon.  In this article I discuss the
influence of the NVO on research into the formation and evolution of low
surface brightness (LSB) galaxies.  In particular, I present the
preliminary results from an NVO-style project that combines the DPOSS and
2MASS datasets to search for giant disk galaxies.

\end{abstract}

\section{The Dataset Revolution}

Astrophysical problems seem to increase in complexity with each successive
generation.  Observationally, new wavelengths and new flux limits bring about
new phenomenon that demands more telescope time and begs theoretical
interpretation.  In the last 15 years we have seen an explosion in the
amount, wavelength coverage and diversity of our datasets that have lead
to numerous discoveries, but have also buried us in the sheer quantity of
information.

Our community has also grown parallel to our data growth, but most of the
high powered observational tools still lie in the possession of a few
institutions.  This disparity in big telescope resources has been offset,
in a large part, by the formation of national data centers and the
distribution of analysis software.  Now an astronomer, regardless of the
size of their home institution, can have access to high quality data and
produce cutting edge science.  With the addition of small grant programs
(i.e. NASA's ADP program), an astronomical industry developed during the
last two decades and discovery has moved from the hands of the few to the
hands of the many.  One only need to compare the impact of HST science on
the astronomical community to that of Keck to see how the existence of
non-proprietary datasets can push forward science.

As datasets have grow larger, there has been a strong emphasis on data
mining and computational power.  However, intelligent and cleverly
designed projects depend more on access to direct tools rather then
sophisticated algorithms.  Thus, many astronomical projects today involve
teams of scientists who bring together the diverse talents needed to
attack the details of massive amounts of data.  This is the goal of NVO,
to provide the infrastructure and intellectual support for astronomical
programs that compliment our already existing observational and theoretical
foundations.

Within the NVO concept there lies the goal of removing the division
between the type of science and the type of data.  Indeed, the greatest
benefit the NVO may bring to us is to eliminate the specialization that
divides our fields of research (stellar versus extragalactic) and between
the wavelength regions (divided mostly due to the technology used to
observe within the spectral regions of interest).  As an example, one area
of astronomy that could derive enormous benefit from the NVO is research
into low surface brightness (LSB) galaxies.

\section{LSB Universe}

One area that is particularly challenging to the observational world is
the universe of LSB objects.  We can only study what we know to exist, and
with respect to galaxies that means the object must reside in some
catalog.  While there has always been a push to find the faintest objects
(meaning the lowest in mass as stellar luminosity maps into baryonic mass)
or the most distant (meaning the closest to the galaxy formation epoch),
it has only been recently that there has been much concern for objects
with low luminosity density.

\begin{figure}
\plotfiddle{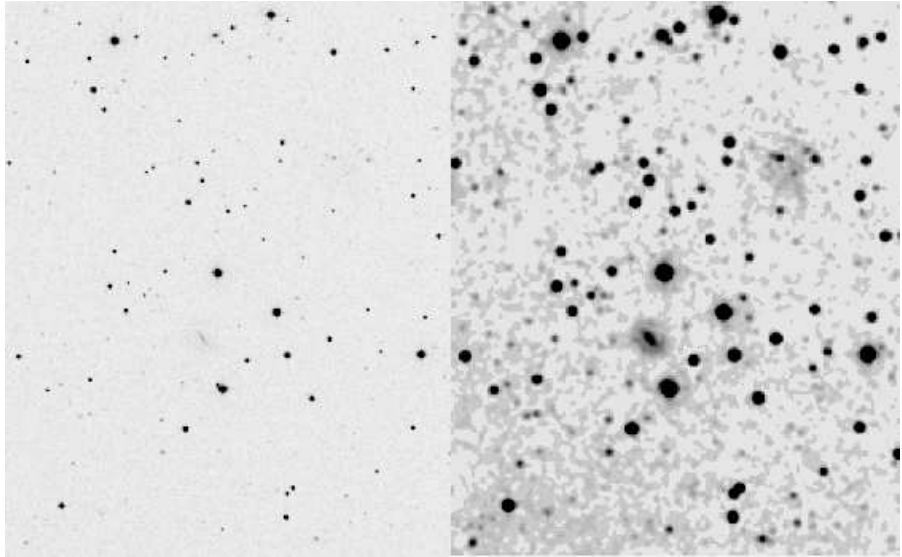}{6.4truecm}{0}{75}{75}{-230}{-200}
\caption{A subset of DPOSS normal (left panel) and smooth with a 5 pixel
gaussian filter (right panel).  A LSB disk galaxy is barely visible in the
normal frame.  A second LSB dwarf galaxy is discovered in the upper
right-hand section of the smoothed frame.  Both objects where detected in
H\,I at Arecibo, the LSB disk has a velocity of 5,550 km/s, the dwarf is
at 2,500 km/s.}
\end{figure}

The LSB realm is interesting to many astrophysical problems.  For example,
star formation is normally a phenomenon associated with high gas density
environments.  Yet, LSB galaxies display many characteristics that indicates
a history of recent star formation.  Thus, studies into their past will
probe star formation in new parameter space.  The range of galaxy
properties requires an examination of the LSB universe because galaxies at
the extreme ends of the mass spectrum (dwarf and giant) tend to be LSB in
nature.  LSB galaxies also differ from their brighter cousins in that
there gas masses often exceed their stellar masses.  Thus, baryon counts of
the Universe at high redshift will be underestimated without some knowledge
of the distribution of LSB galaxies.

In some sense, the lack of pursuit of LSB research is technical in nature.
To find the faintest, or most distant, galaxies becomes a simple process
of building larger and larger collecting surfaces.  However, achieving
fainter levels of surface brightness to explore the LSB universe battles
against the natural glow of the night sky and is not overcome with larger
pieces of glass.  Space imaging has the immediate advantage of getting
above the sky glow, but the emphasis in space has been on small and faint,
so pixel sizes to take advantage of high resolution images work against
LSB objects by reducing the number of counts per pixel.  Even in the
non-optical portions of the spectrum the emphasis is always first on the
detection threshold of point sources rather than design for sky brightness
(see, for example, the 2MASS survey).

The first expeditions into the LSB universe where taken, of course, by
Zwicky who hypothesized on the existence of `hidden' galaxies as a counter
to Hubble's notion that the galaxy luminosity function is gaussian in
shape.  By the 1960's, numerous galaxy catalogs by Arp, Sandage, van den
Bergh (DDO) and de Vaucouleurs (RC2) had defined the Hubble classification
system.  These catalogs were primarily defined by HSB galaxies, but there
were always appendices or notes concerning `diffuse' objects, usually
assumed to be nearby dwarfs.

Disney (1976) was the first to place the concept of galaxy visibility in
an analytic form and to demonstrate that the mean central surface
brightness of our galaxy catalogs was, in fact, a function of the natural
sky brightness and not imposed by astrophysics.  While the importance of
this work is recognized today, galaxy evolution was a relatively new field
at the time and the study of LSB galaxies remained in the background (no
pun intended) until the mid-1980's.

\section{LSB Detection}

There was very little that the typical observational astronomer could do
about the night sky problem until the mid-1980's with the evident of the
Second Palomar Sky Survey.  While the sky brightness had only degraded
since the first Sky Survey, the finer emulsions and deeper plate material
allowed for, at least, a cursory examination of the difference between the
angular limited UGC and the new plate material.  This resulted in the
PSS-II LSB catalogs (Schombert \& Bothun 1988, Schombert \etal 1992,
Schombert \etal 1997) which were visual surveys, but demonstrated an
increase of one mag arcsec$^{-2}$ to the old catalogs.  More importantly,
it provided a jumpstart to the LSB field by simply providing a new list of
objects in which to study.

The first visual surveys were extremely crude but spurred a more exacting
search for LSB's using CCDs (O'Neil, Bothun \& Cornell 1997, Dalcanton
\etal 1997).  Flattening is always a key parameter for LSB galaxy
detection.  Most CCDs on 1-meter class telescopes are sky limited in a few
minutes of exposure time.  The critical component in finding LSB galaxies,
and measuring fluxes, is how well you know the sky value and how flat (on
large and small scales) you can make your data.  Transit CCD surveys offer
the best method for sky flattening, since they allow the sky to pass
through each pixel which is then summed for the entire scan.  Long scans
can suffer from temporal variations, but these can at least be quantified.

LSB detection is functionally a very difficult problem.  The standard
procedure is to detect all the HSB objects, mask them out, smooth and run
the detection scheme again.  Unfortunately, this will usually remove any
LSB galaxies with bright bulges.  They will be in the original catalog,
but their LSB nature may not be known from whatever parameters are stored
in the catalog.

\begin{figure}
\plotfiddle{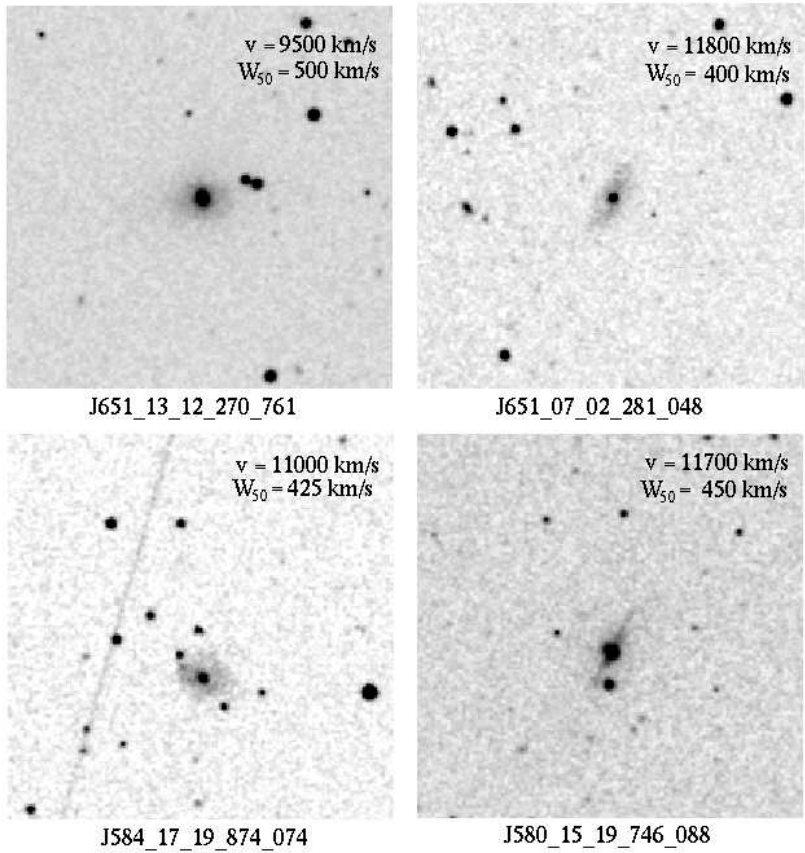}{11.5truecm}{0}{75}{75}{-235}{-250}
\caption{Four AGN Malin objects discovered from a combined search of DPOSS
and 2MASS datasets.  The near-IR 2MASS catalogs provide the low luminosity AGN
sources (due to their anomalous colors) typical to the Malin class.  The optical catalog is fast filtered for the presence of an LSB disk.}
\end{figure}

One example of a smooth and search algorithm is shown in Figure 1.  Here a
sample of the sky from a J plate of DPOSS is shown in its raw form.  A
single LSB disk galaxy is obvious near the center of the frame.  Smoothing
with a gaussian filter enhances the detectable of the LSB disk and also
discovers a second LSB galaxy in the top right portion of the scan.
Both objects were detected in H\,I at Arecibo with velocities of 5,550
km/s and 2,500 km/s respectfully.

The smoothed image demonstrates the two major difficulties in automatizing
LSB detection algorithms.  The first is that the number of pixels distorted by
stellar sources is extremely high at the surface brightness levels of
interest.  It is practically impossible to detect, in an automatic
fashion, LSB objects near stars (although the human brain seems to carry
out the task fairly well).  Second, the uncorrelated background noise
varys substantially over even this small piece of sky.  This makes a
systematic survey, to specific threshold levels, a technical challenge (what NASA
would call completeness and reliability).

\section{AGN Malin Search}

One of the most intriguing classes of LSB galaxies is the supergiant disk
systems, the Malin class.  The prototype to this class, F568-6, was
discovered from a visual search of the PSS-II (Bothun \etal 1990).  While
low in central surface brightness ($\mu_o = 23.4 B$ mag arcsecs$^{-2}$),
F568-6 is by no means low in luminosity ($M_B = -21.1$) nor low in total
or H\,I mass.  The Malin class contains the largest galaxies in the
Universe, yet are notorious difficult to find and catalog (Sprayberry,
Impey \& Irwin 1996).

One of their properties provides a promising avenue for the cataloging of
a significant number.  Most Malin class galaxies have a weak AGN in their
core.  The current theory is that the copious gas supply in the disk
provides the fuel for a central engine, even if at a low intensity
(Schombert 1998).  While the weak AGN appears as a point source, its
near-IR colors would distinguish it from a stellar SED.  Unfortunately, a
survey of the sky in the near-IR will not, by itself, identify the Malin
objects since the sky brightness at 2.2 $\mu$m is 2000 times higher than
in the optical making their disk regions invisible.  Thus, this project
requires a `virtual observatory', in this case the combination of two
existing databases, DPOSS (optical) and 2MASS (near-IR).

The procedure is straight-forward, first isolate all the objects in the
2MASS catalog with non-thermal colors (i.e. outside some boundary defined
by normal stars).  Second, search the near-IR source positions on the blue
plates of DPOSS with a fast area scan.  A series of circular apertures are
placed around the point source then tested against the local sky.  LSB
galaxy detection is effective if only a particular region is being tested
against the background since varying diameters are checked which maximize
the signal from the LSB disk versus the signal from sky.

A preliminary search was undertaken last winter using eight plates from
DPOSS that contained some fraction of 2MASS coverage (about 3 square
degrees).  Forty candidates were produced of which ten of these objects
were searched with the new, upgraded Arecibo telescope.  The Gregorian
system at Arecibo has a much wider velocity range, a critical element
since the large Malin objects tend to be at velocities greater than 8,000
km/s.  Eight of the ten candidates were detected at 21-cm.

Four of the detection's DPOSS images are shown in Figure 2.  All have the
characteristics AGN nucleus surrounded by a LSB disk.  All eight also have
H\,I widths in excess of 350 km/s (the typical spiral has a rotation width
of 250 km/s).  Assumingly these galaxies follow the baryonic TF relation
(McGaugh \etal 2000), then their masses will exceed $10^{12} M_{\sun}$.

\section{Conclusions}

It is universally recognized that the NVO would be a powerful tool for the
astronomical community.  A particular emphasis will be placed on the need
for the NVO to make the most efficient use of the large datasets in our
present holdings and to future projects.  However, perhaps one of the most
common uses of the NVO will be of the one human/one workstation type
project.

One such project I have described herein, the search for AGN Malin
galaxies typifies my vision of how small projects will use the NVO.
The merger of multi-wavelength datasets and simple tools, combined
with a researchers experience in the astrophysical phenomenon, was used
here to achieve a catalog of a new, and exciting, type of galaxy.

While each of our individual research interests may reap the benefits of
the NVO, there is no doubt that the sum of the contributions of many small
projects will also service to build the framework of the NVO.  A system
that we hope will be wavelength and research field independent.

\acknowledgements

I wish to thank my collaborators in the LSB universe (G. Bothun, J. Eder,
S. McGaugh and K. O'Neil) for their insights.  Many different telescopes
have contributed to the LSB searches, but a special thanks is given to the
Arecibo scientists and staff for their speedy upgrade which has increased
LSB's visibility by many orders of magnitude.  I also wish to thank the
organizers of the conference for inviting me and the NASA grant
(NAG5-6109) which has supported my LSB programs.

\end{document}